\documentclass[pra,
                preprint,
                amsmath,
                amssymb,
                showpacs,
                superscriptaddress]{revtex4-1}
\usepackage{graphicx} 
\usepackage{dcolumn}  
\usepackage{bm}       
\usepackage{amsfonts}
\usepackage{dsfont}
\usepackage{csquotes}

\usepackage{ulem}
\usepackage{color}

\usepackage{float}
\usepackage{physics}

\begin{document}

\title{Probing phonon-driven symmetry alterations in graphene via high-harmonic spectroscopy}

\author{Navdeep Rana}
\affiliation{%
Department of Physics, Indian Institute of Technology Bombay,
            Powai, Mumbai 400076  India}
            
\author{Gopal Dixit}
\email[]{gdixit@phy.iitb.ac.in}
\affiliation{%
Department of Physics, Indian Institute of Technology Bombay,
            Powai, Mumbai 400076  India}

\date{\today}


\begin{abstract}
High-harmonic spectroscopy has become an essential ingredient  in probing 
various ultrafast electronic processes in 
solids with sub-cycle temporal resolution. 
Despite its  immense importance, sensitivity of  
high-harmonic spectroscopy to phonon dynamics in solids is not well known.  
This work addresses this critical question and 
demonstrates the potential of high-harmonic spectroscopy in probing intertwined phonon-electron dynamics in 
solids. A pump pulse excites in-plane optical phonon modes in monolayer graphene and 
a circularly polarised pulse is 
employed to probe the excited phonon dynamics that generates higher-order harmonics. 
We show that the coherent phonon dynamics alters the dynamical symmetry  of graphene with 
the probe pulse and leads the generations of the  symmetry-forbidden harmonics. 
Moreover, sidebands associated with the prominent harmonic peaks are generated as a result of the coherent dynamics. It is found that the 
symmetries and the characteristic timescale of the excited phonon mode determine the polarisation and 
positions of these sidebands. Present work opens an avenue in time-resolved probing of 
phonon-driven processes and dynamical symmetries in solids with sub-cycle temporal resolution.
\end{abstract}

\maketitle 

\section{Introduction}
Vibrations of atoms within molecules and solids are fundamental processes 
that regulate several physical, optical and chemical properties of matter. 
When light triggers atomic vibrations, atoms exhibit periodic oscillations in a particular fashion
and these light-induced vibrations could potentially lead to the modifications in various symmetries of solids. 
These modifications are dynamic in nature and result in several transient phenomena, such as light-induced superconductivity~\cite{hu2014optically, mitrano2016possible}, 
vibrationally-induced magnetism~\cite{rini2007control, nova2017effective}, 
and switching of electrical polarisation~\cite{mankowsky2017ultrafast}, to name but a few. 
Thus, time-resolved mapping of the interplay of lattice vibration with electronic motion on electronic timescale 
is essential to comprehend several ubiquitous phenomena in solids, such as structural phase transition~\cite{bansal2020magnetically, hase2015femtosecond}, thermal~\cite{Niedziela_2019, Bansal_2018}
and optical properties~\cite{katsuki2013all, Fultz2010, gambetta2006real}; and predicting new concepts in solids.
Several  spectroscopy and imaging-based methods are employed to probe lattice vibrations in solids ~\cite{dhar1994time, debnath2021coherent, graf2007spatially, virga2019coherent, koivistoinen2017time, rana2021four, brown2019direct, flannigan2018electrons}. 
However, probing transiently-evolving intertwined lattice-electronic dynamics and 
dynamical symmetries of solids in the presence of light in a single experimental setup are challenging. 
Present work addresses this crucial problem.  

High-harmonic generation (HHG) is a non-perturbative nonlinear frequency up-conversion process and 
is sensitive to the sub-cycle electron dynamics driven by intense laser. 
Over the last decade,  high-harmonic spectroscopy became an emerging method to interrogate various 
equilibrium  and non-equilibrium  properties of solids by investigating the emitted spectrum during HHG~\cite{luu2015extreme, schubert2014sub, mrudul2021light, mrudul2021controlling, hohenleutner2015real, zaks2012experimental, pattanayak2020influence, langer2018lightwave, mrudul2020high, luu2018measurement, banks2017dynamical, pattanayak2019direct, bai2020high, imai2020high, borsch2020super}. 
In spite of the tremendous applications of  high-harmonic spectroscopy,  
impact of lattice vibration on  HHG from solids remains uncharted territory, except recent work~\cite{ginsberg2021optically, neufeld2022probing}.  
Present work focuses to highlight  the abilities  of high-harmonic spectroscopy in time-resolved mapping of 
the interplay of coherent lattice vibrations and electronic motions; and 
transiently-evolving symmetries of solids during the dynamics.
 
In the following, we will demonstrate  that the high-harmonic spectroscopy of 
the coherent lattice dynamics leads to the generations of the higher-order sidebands along with 
the main harmonic peaks in the high-harmonic spectrum. 
The frequency and symmetry of the coherently excited phonon mode are 
imprinted  in the position and polarisation of the sidebands, respectively.   
Moreover, symmetry-forbidden harmonics are allowed due to the symmetry alterations caused by the
coherent lattice dynamics. 

To illustrates the sensitivity of the coherent phonon dynamics to high-harmonic spectroscopy, 
two-dimensional graphene with 
 $\textbf{D}_{6\textrm{h}}$ point group symmetry is chosen. 
The phonon spectrum of graphene is consist of three acoustic  and three  optical phonon branches. 
Out of the three optical phonon modes, one phonon mode is out-of-plane 
where the vibrations are out of the two-dimensional plane of graphene, 
and other two are in-plane modes in which  lattice vibrations are confined within the plane of  graphene~\cite{kim2013coherent}.  
In-plane optical phonon modes are considered in this work.  
At the $\mathsf{\Gamma}$ point,  there are 
two degenerate in-plane optical modes with 
phonon frequency of 194 meV, which corresponds to phonon oscillation period of $\sim$ 21 femtoseconds~\cite{kim2013coherent}. 
These phonon modes are represented as E$_{2\textrm{g}}$ or $\mathsf{G}$ modes and are Raman active. 
It is known that light can couple to a phonon mode 
at the Brillouin zone centre and therefore E$_{2\textrm{g}}$ Raman-active mode can be excited by stimulated  Raman excitation either via a  broad pulse that covers a bandwidth of 194 meV or  via using  two laser pulses with a difference of 194 meV in photon energy. 
By tuning the polarisation of the phonon exciting pulse either along   
$\mathsf{\Gamma-K}$ or $\mathsf{\Gamma-M}$ direction, one of the two in-plane phonon modes can be excited selectively.

\section{Computational Methods}
To incorporate coherent phonon dynamics,  nearest-neighbour tight-binding Hamiltonian is extended 
from static to time domain as
\begin{equation}
\mathcal{\hat{H}}_\textbf{k}(t) =  - \gamma(t)\sum_{i\in nn} e^{i\textbf{k}\cdot \textbf{d}_i(t)} \hat{a}_\textbf{k}^{\dagger} \hat{b}_\textbf{k} + \textrm{H. c.}
\label{eq:tth}
\end{equation}
Here, $\gamma(t)$ = $\gamma_0~e^{-(|\textbf{d}_{i}(t)|-a)/\delta}$ 
is the time-dependent nearest-neighbour hopping energy, which is  modelled to capture  the temporal variations in the relative distance 
between nearest-neighbour atoms ($\textbf{d}_i = a = 1.42 $~\AA)~\cite{mohanty2019lazy}. 
$\gamma_0 =$  2.7 eV is the nearest-neighbour hopping energy and $\delta$ = 0.184a$_0$ is the 
width of the decay function with a$_0$ = 2.46~\AA~as the lattice parameter of the equilibrium structure~\cite{moon2013optical}.

Semiconductor-Bloch equations corresponding to time-dependent Hamiltonian given in Eq.~(\ref{eq:tth}) are
solved as 
\begin{subequations}
\begin{align}
    \frac{d}{dt}\rho_{cv}^{\textbf{k}} &= \left[-i\varepsilon_{cv}(\textbf{k}_t, t)+\frac{1}{\textrm{T}_2}\right]\rho_{cv}^{\textbf{k}} + i\textbf{E}(t)\cdot\textbf{d}_{cv}(\textbf{k}_t, t)\left[\rho_{vv}^{\textbf{k}}-\rho_{cc}^{\textbf{k}}\right]~~\textrm{and} \\
    \frac{d}{dt}\rho_{vv}^{\textbf{k}} &= i\textbf{E}(t)\cdot \textbf{d}_{vc}(\textbf{k}_t, t)\rho_{cv}^{\textbf{k}} + \textrm{c.c.} \end{align}
\label{SBE}
\end{subequations} 
Here, the vector potential  and the electric field corresponding to the driving laser field are represented as 
$\textbf{A}(t)$ and 
$\textbf{E}(t)$, respectively, and are related as  $\textbf{E}(t)$ = $-d\textbf{A}(t)/dt$. 
$\textbf{d}_{cv}(\textbf{k})$ and $\varepsilon_{cv}(\textbf{k})$  are, respectively, the dipole matrix elements  
and the band-gap energy   between conduction and valence  bands at given $\textbf{k}$  with $\textbf{d}_{cv}(\textbf{k}) = i\langle c,\textbf{k} |\nabla_\textbf{k}|v,\textbf{k}\rangle$. 
The matrix elements become time-dependent due to coherent phonon dynamics and are updated smoothly
at each consecutive time-steps during the phonon dynamics.  
To account the interband decoherence, 
a phenomenological term  with a constant dephasing time $\textrm{T}_2$ is introduced. 
Here, $\textbf{k}_t = \textbf{k} + \textbf{A}(t)$. 

High-harmonic spectrum is calculated  as
\begin{equation}
\mathcal{I}(\omega) = \left| \mathcal{FT} \left(  \dfrac{\rm{d}}{{\rm d}t} \textbf{J}(t)    \right) \right|^2, 
\end{equation}
where $\mathcal{FT}$ stands for the Fourier transform. The total current \textbf{J}(t) is 
calculated by integrating $\textbf{J}(\textbf{k}, t)$ over 
the entire Brillouin zone as 
\begin{equation}
\textbf{J}(\textbf{k}, t)   = \sum_{m,n  \in \{c,v\} } \rho_{mn} ^{\textbf{k}} (t)  \textbf{p}_{nm}(\textbf{k}_t, t), 
\end{equation}
where $\textbf{p}_{nm}(\textbf{k}) = \langle n,\textbf{k}|\nabla_\textbf{\textbf{k}}\mathcal{\hat{H}}_\textbf{k}| m,\textbf{k}\rangle$ is momentum matrix-element. 
We have adopted the same methodology for HHG from a solid with phonon dynamics as given in our earlier work~\cite{rana2022high}.
The ellipticity and the phases of harmonics are estimated by following the recipe given in Ref.~\cite{tancogne2017ellipticity}. 

It is assumed that a pump-pulse initiate the coherent excitation of an in-plane Raman active phonon mode in graphene. The excited phonon mode is probed by the high-harmonic generating pulse. 
The excitation of the phonon mode is approximated by direct initiation of the coherent vibrations of carbon 
atoms  within  adiabatic approximation. 
Circularly-polarised pulse with a  wavelength of 2.0 $\mu$m and peak intensity of 
1$ \times 10^{11} $ W/cm$^2$ is used to generate high-harmonics in graphene with and without coherent phonon dynamics. 
The pulse duration is 100 fs, which is  much longer than an oscillation period of in-plane phonon dynamics 
$\simeq$ 21 fs. Moreover, the coherence times of these phonon modes of around 1 ps are 
reported~\cite{jeong2015coherent, hubener2018phonon}. 
Parameters of the harmonic generating laser are similar to the ones used earlier for probing  electron dynamics in graphene~\cite{heide2018coherent,  yoshikawa2017high}. 

\section{Results and Discussion}

\begin{figure}
\includegraphics[width=  \linewidth]{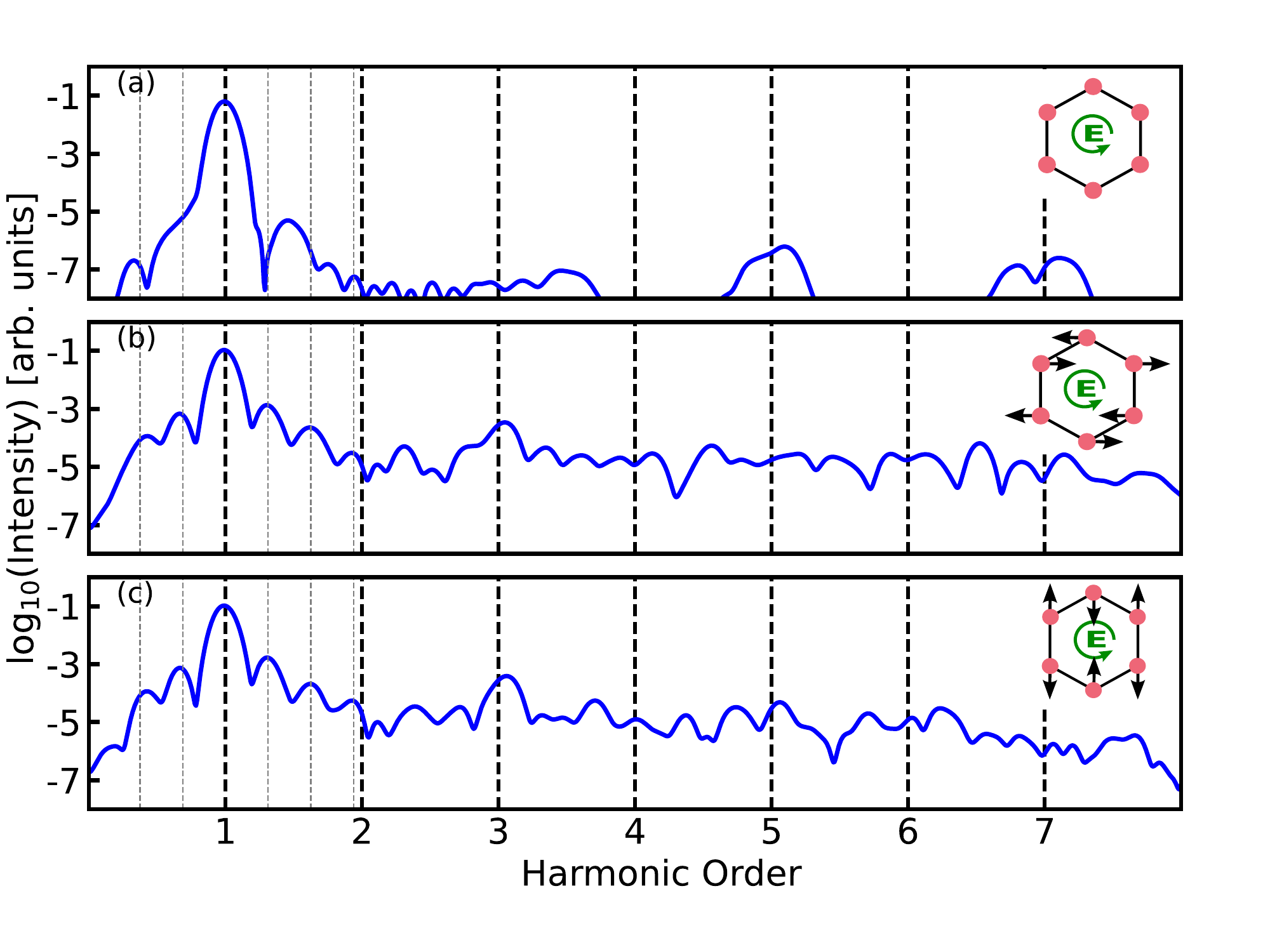}
\caption{High-harmonic spectra, generated by left-handed 
circularly polarised laser pulse, of two-dimensional  graphene with 
and without coherent phonon dynamics.  (a)  The spectra of the graphene without  phonon dynamics. 
The spectra of graphene with the coherent in-plane  (b) longitudinal optical ($\textsf{iLO}$) phonon mode  and
(c) transverse optical ($\textsf{iTO}$) phonon mode. 
In all the spectra, sidebands corresponding to the prominent harmonic peaks are identified at 
frequencies ($\omega_{0} \pm  n \omega_{\textrm{ph}}$) with $\omega_{\textrm{ph}}$ as the phonon frequency, 
$\omega_0$  as the frequency of the probe laser pulse and 
$n$ is  an integer. 
The unit cell of the graphene with the eigenvector of a particular phonon mode and polarisation of the harmonic generating probe pulse are shown in the respective insets.
Results are presented for $\textrm{T}_2 = $  10 fs and 
a maximum 0.03a$_0$ displacement of  atoms from their 
equilibrium positions during coherent phonon dynamics where a$_0$ is 
the lattice parameter of the equilibrium structure.
Our findings remain qualitatively same  for $\textrm{T}_2$ ranging from 5 to 30 fs and 
for  displacements ranging from 0.01a$_0$ to  0.05a$_0$ with respect to the equilibrium positions of the carbon atoms.}
\label{spectra}
\end{figure}

High-harmonic spectrum corresponding to monolayer graphene without phonon dynamics 
is presented in Fig.~\ref{spectra}(a).  
We have employed left-handed circularly polarised laser pulse for HHG in graphene with and without phonon dynamics.
As dictated  by the symmetry constraints and selections rules, 
it is expected that circularly polarised  pulse
yields $(6m \pm 1)$-orders of harmonics from an inversion-symmetric graphene with the six-fold symmetry~\cite{alon1998selection, chen2019circularly}. Here,  $m = 0, 1, 2, \ldots$ is a positive integer. 
In this case,  third harmonic is symmetry forbidden. 
On the other hand, linearly polarised laser pulse leads to  $(2m + 1)$-orders of harmonics as shown earlier~\cite{mrudul2021high}.  
Our results shown in Fig.~\ref{spectra}(a)  are consistent with the selection rules and earlier report~\cite{alon1998selection, chen2019circularly}. 
Graphene is not chiral in nature, so left- and right-handed 
circular laser pulses yield same harmonic spectra. 

After discussing HHG from graphene without phonon dynamics, 
let us investigate how the in-plane phonon modes 
affect the harmonic spectrum shown in  Fig.~\ref{spectra}(a). 
For this purpose, we coherently excite one of the  two degenerate   in-plane phonon modes
and assume that the excitation is done prior to the probe harmonic pulse. 
Figure~\ref{spectra}(b) presents the harmonic spectrum  corresponding to 
coherently excited longitudinal optical ($\textsf{iLO}$) phonon mode.  
The spectrum in Fig.~\ref{spectra}(b)
is drastically different from the one without phonon dynamics [see Fig.~\ref{spectra}(a)]. 
There are only odd harmonics in the spectrum as the $\textsf{iLO}$ phonon
mode preserves the inversion symmetry in graphene~\cite{rana2022high}. 
Moreover, the spectrum exhibits multiple sidebands along with the main odd harmonics as evident from Fig.~\ref{spectra}(b). 
The coherent excitation of the in-plane transverse optical ($\textsf{iTO}$) phonon mode also leads to  
multiple sidebands along with the  odd harmonics as visible from Fig.~\ref{spectra}(c). 
In both cases, the energy separation between the successive sidebands is equal to the energy of the  
excited phonon  ($\textsf{iLO}$  or $\textsf{iTO}$) mode, i.e.,  194 meV. 
Therefore, the energy of the excited phonon mode is encoded in the spectra. 
Apparently, it seems that the spectra is insensitive to the symmetry of the excited phonon mode as  
both $\textsf{iLO}$  and $\textsf{iTO}$ phonon modes yields similar harmonic spectra [see Figs.~\ref{spectra}(b) and ~\ref{spectra}(c)]. 
In the following, we will show that this is not the case and the symmetry of the excited phonon mode 
is encoded in 
the polarisation properties of the spectra. 

Not only coherent phonon dynamics leads to the generation of the  multiple sidebands   
but also the forbidden harmonics become allowed. 
As stated earlier, third harmonic is absent for  the circular laser driven HHG from 
graphene without phonon [see Fig.~\ref{spectra}(a)].
However, dynamics of the coherent E$_{2\textrm{g}}$ 
phonon  mode reduces graphene's six-fold symmetry into two-fold dynamically, which 
allows the generation of $(2m \pm 1)$ harmonic orders.  
The presence of  the third harmonic in both cases, graphene with ($\textsf{iLO}$  or $\textsf{iTO}$) phonon mode,
is a signature of the symmetry reduction as evident from Fig.~\ref{spectra}(b) and ~\ref{spectra}(c). 
At a glance, it seems that the criteria for HHG  is same for 
linearly polarised laser pulse  [$(2m + 1)$ orders and third harmonic]  and the combination of the phonon-driven symmetry reduction  with circularly polarised laser pulse [$(2m \pm 1)$ orders and third harmonic].  
To distinguish the two situations, let us analyse the  polarisation properties of the emitted harmonics.
 
\begin{figure}
\includegraphics[width=\linewidth]{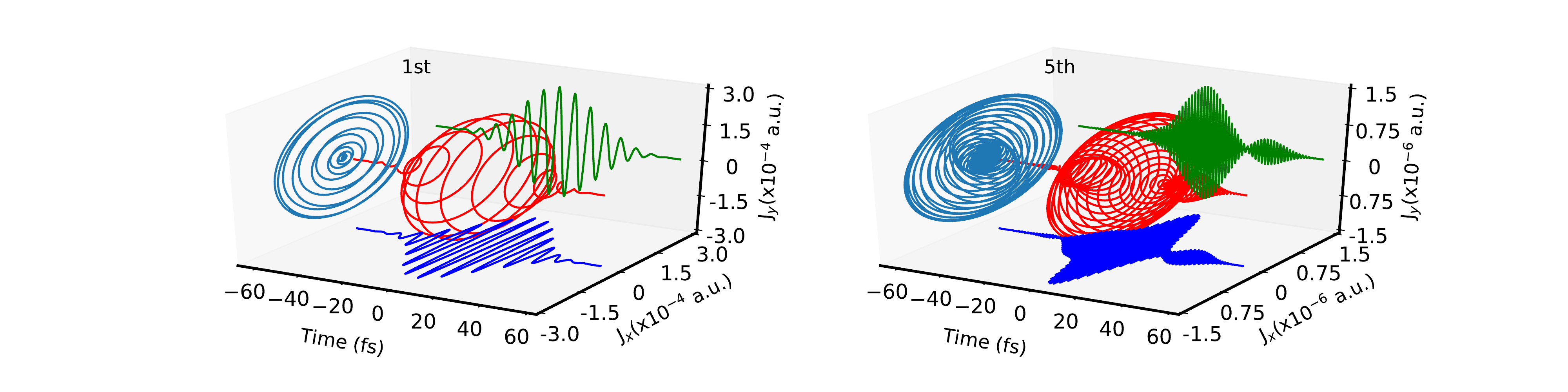}
\caption{Lissajous figure (in cyan), total (in red), $x$ (in blue) and $y$  (in green) 
components of the first and  the fifth harmonics in time-domain corresponding to 
the spectrum shown in Fig.~\ref{spectra}(a) for  graphene without phonon dynamics.}
\label{polarization_no_latt}
\end{figure}

Figure~\ref{polarization_no_latt} displays the time-domain  representations of the $x$ and $y$ components 
of the  first and fifth harmonics of the spectrum in  Fig.~\ref{spectra}(a).  
It is known that the polarisation of a given harmonic for a material with $l$-fold symmetry is
determined by $lm  + \sigma$, where 
$\sigma =  + (-)1$ represents the $m^{\textrm{th}}$ harmonic's  polarisation, 
which is  same (opposite) as the helicity of the driving laser pulse~\cite{alon1998selection}. 
It is straightforward to see that  $(lm - 1)^{\textrm{th}}$ and $(lm + 1)^{\textrm{th}}$ harmonics are circularly polarised with $\sigma = -1$ and $\sigma = 1$, respectively. 
In present case,  the first and fifth harmonics are circularly  polarised  
with opposite helicity and is consistent with  earlier findings~\cite{saito2017observation, chen2019circularly}.

\begin{figure}
\includegraphics[width=\linewidth]{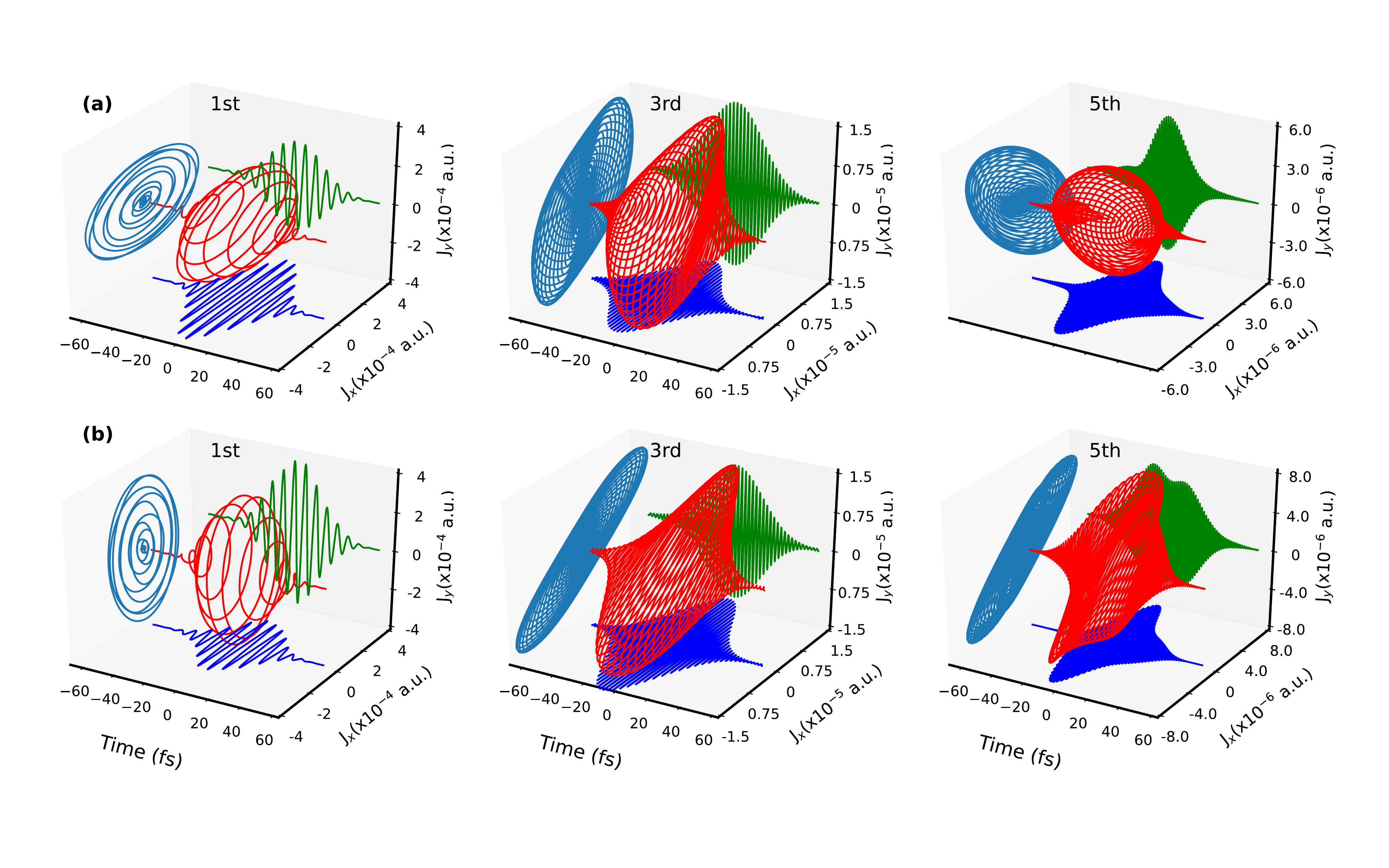}
\caption{Same as Fig.~\ref{polarization_no_latt} for the first, third and  
fifth harmonics in time-domain for graphene with (a) $\textsf{iLO}$ and (b) $\textsf{iTO}$ phonon modes. 
In the case of $\textsf{iLO}$ ($\textsf{iTO}$) phonon mode, the ellipticities of the first, third and  
fifth harmonics are 0.63 (0.62), 0.77 (0.79), and 0.93 (0.96), respectively. 
Also, the phase differences between $x$ and $y$  components  of the first, third and  
fifth harmonics are  90$^{\circ}$ (90$^{\circ}$), 40$^{\circ}$ (35$^{\circ}$), and 125$^{\circ}$ (45$^{\circ}$), respectively.}
\label{polarization_latt}
\end{figure}

As stated above, $(2m \pm 1)$ harmonic orders are allowed 
due to phonon-driven dynamical symmetry reduction from six-fold to two-fold, which 
leads to the generation of the third harmonic. 
Moreover, this dynamical symmetry reduction also alters the 
polarisation properties of the emitted harmonics. 
The $x$ and $y$ components 
of the first, third, and fifth harmonics in  time domain for graphene 
with $\textsf{iLO}$  and $\textsf{iTO}$ phonon modes are presented in 
Figs.~\ref{polarization_latt}(a) and  ~\ref{polarization_latt}(b), respectively. 
As evident from the figure, the ellipticity of the first harmonic reduces drastically from 1 for graphene without phonon to 0.63  for graphene with phonon. 
The change in the ellipticity can be understood as follow: 
When $\textsf{iLO}$ phonon mode is excited, carbon atoms vibrate along the \textsf{X} direction, which
increases the velocity of the  electrons in the \textsf{X} direction. 
It is known that the intraband current is proportional to the velocity,  
and low-order harmonics in graphene are dominated by the intraband current~\cite{mrudul2021high, vampa2015semiclassical}. 
Thus, the major axis of the ellipse is along the \textsf{X} direction in the case of  the first harmonic, which
reduces the ellipticity from 1 to 0.63.

Similarly,  the excitation of the $\textsf{iTO}$ mode leads  the vibrations of the atoms along  \textsf{Y} direction. 
This provides an additional velocity component to electrons in the \textsf{Y} direction, which translate to 
the major axis of the ellipse  along the \textsf{Y} direction. 
The ellipticity of the fifth harmonic, corresponding to graphene without phonon to with 
$\textsf{iLO}$ ($\textsf{iTO}$) phonon mode, changes significantly, i.e., from 1  
to 0.93 (0.96).  
Not only the phonon dynamics modifies the ellipticity of the harmonics  significantly  but also 
changes  the phase between the $x$ and $y$ components of the harmonics. 
In the case of $\textsf{iLO}$ ($\textsf{iTO}$) mode, the phase differences between the components for 
first and fifth harmonics are 90$^{\circ}$ (90$^{\circ}$) and  125$^{\circ}$ (45$^{\circ}$), respectively (see  Fig.~\ref{polarization_latt}).  
The ellipticity and the phase difference of the third harmonic for graphene with $\textsf{iLO}$ ($\textsf{iTO}$) mode is 0.77 (0.79), and 40$^{\circ}$ (35$^{\circ}$), respectively. 
Thus, the changes in the ellipticity and phase indicate that the harmonics are sensitive to the symmetry of the excited phonon mode as reflected from Fig.~\ref{polarization_latt}. 

\begin{figure}
\includegraphics[width=\linewidth]{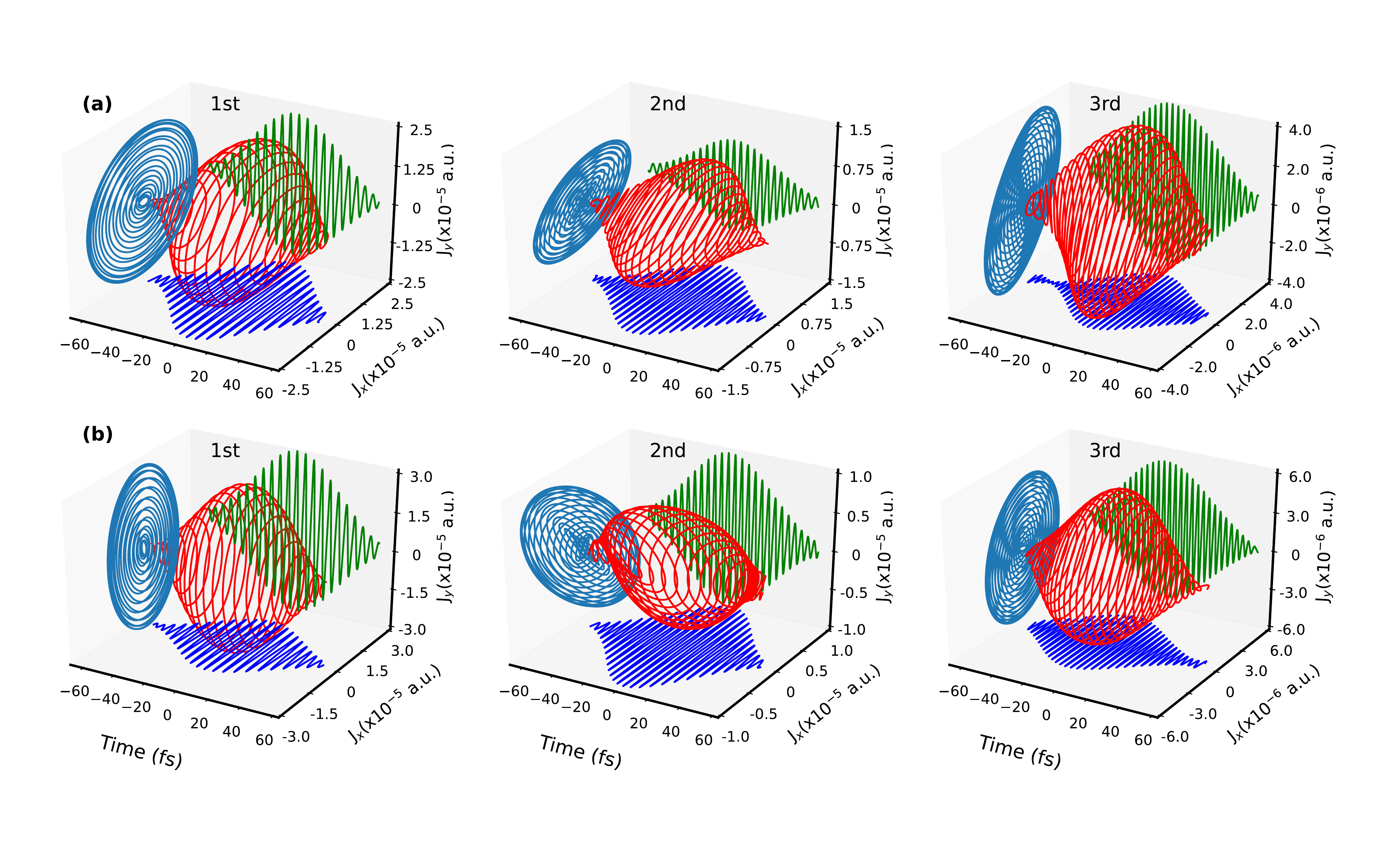}
\caption{Same as Fig.~\ref{polarization_no_latt} for the first, second and third sidebands in time-domain 
associated with the 
first harmonic  for graphene with (a) $\textsf{iLO}$ phonon mode corresponding to Fig.~\ref{spectra}(b), 
and (b) $\textsf{iTO}$ phonon mode corresponding to Fig.~\ref{spectra}(c).  
The ellipticities of the first, second and third sidebands corresponding to $\textsf{iLO}$  ($\textsf{iTO}$)  
phonon mode  are  0.98 (0.56), 0.72 (0.95), and 0.59 (0.67), respectively. 
The phase differences between $x$ and $y$  components  of the first, second and third	
sidebands corresponding to $\textsf{iLO}$  ($\textsf{iTO}$)  
phonon mode are 85$^{\circ}$ (90$^{\circ}$), 70$^{\circ}$ (135$^{\circ}$), and 60$^{\circ}$ (75$^{\circ}$), respectively.}
\label{polarization_side}
\end{figure}

After demonstrating how the information of the excited phonon mode and its symmetry are 
imprinted  in the  main harmonics and their polarisation properties, 
let us analyse what informations are encoded in the sidebands associated with prominent  harmonics.  
Time-domain picture of the $x$ and $y$ components 
of the first, second and third sidebands corresponding to the first harmonic corresponding to  graphene with $\textsf{iLO}$ phonon mode is shown in Fig.~\ref{polarization_side}(a). 
All three sidebands have nonzero $x$ and $y$ components as evident from the figure.
The same is true for the sidebands associated with  $\textsf{iTO}$ phonon excitation  as reflected  from Fig.~\ref{polarization_side}(b). 
Moreover, the ellipticities  of the first, second and third sidebands of the first harmonic 
corresponding to graphene with $\textsf{iLO}$  ($\textsf{iTO}$)  phonon mode read as  
0.98 (0.56), 0.72 (0.95), and 0.59 (0.67), respectively. 
Also,  the phases between the  $x$ and $y$ components of the first, second and third sidebands 
of the  $\textsf{iLO}$  ($\textsf{iTO}$)  mode
are 85$^{\circ}$ (90$^{\circ}$), 70$^{\circ}$ (135$^{\circ}$), and 60$^{\circ}$ (75$^{\circ}$), respectively. 
Thus, the analysis of Fig.~\ref{polarization_side} establishes that the polarisation and the phase properties of the sidebands are different for different phonon mode. 
However, it is not obvious why the sidebands have nonzero $x$ and $y$ components,  
whereas a particular phonon mode ($\textsf{iLO}$ or $\textsf{iTO}$)  induces  atomic vibrations 
along  a particular direction (\textsf{X} or \textsf{Y}). 

To know the origin of the nonzero $x$ and $y$ components of the sidebands, we employe Floquet formalism  
to graphene with  coherently excited phonon mode and  circularly polarised probe pulse. 
Floquet formalism   determines 
dynamical (spatio-temporal) symmetries  (DSs) of the system, which dictate selection rules  for the sidebands.
The $n^{\rm th}$-order sideband obeys the symmetry constraint  as 
$\hat{X}^t\textbf{E}_{s,n}(t) [\hat{X}^t\textbf{E}(t)]^\dagger$ = $\textbf{E}_{s,n} \textbf{E}^\dagger(t)$
with the condition that the spatial symmetries of $\hat{X}^t$ and the probe pulse are same~\cite{nagai2020dynamical}. Here, $\hat{X}^t$ is a DS, and 
the electric fields associated with 
$n^{\rm th}$-order sideband and the probe laser are represented by 
$\textbf{E}_{s,n}(t)$ and $\textbf{E}(t)$,  respectively. 
The Raman tensor, denoted as 
$\mathcal{R}_n(t) = \textbf{E}_{s,n}(t)\textbf{E}(t)^\dagger$,  has to be
invariance under the operation of the DSs~\cite{nagai2020dynamical}. 
In the following, we will follow the same treatment  as given in Refs.~\cite{nagai2020dynamical, rana2022high} 
to investigate  the properties of the sidebands using  DSs within Floquet formalism.

In the present case,  $\mathcal{D}_1 = \hat{\sigma}_y \cdot \mathcal{T}$ is the DS, 
which leaves the system with $\textsf{iLO}$ phonon mode invariant.  Here, 
$\hat{\sigma}_{y}$ is the reflection with respect to 
$y$-axis, and $\hat{\mathcal{T}}$ is the time-reversal operator. 
Thus, the condition associated with the sidebands is determined as 
$\hat{\mathcal{D}_1}\mathcal{R}_n(t) = \mathcal{R}_n(t)$ with 
\begin{equation}
\mathcal{R}_n(t) = \mathbf{E}_{s,n}(t)\mathbf{E}(t)^\dagger
= \begin{bmatrix}
	E_{s,n_{x}}E^{*}_{x} & E_{s,n_{x}}E^{*}_{y} \\
	E_{s,n_{y}}E^{*}_{x} & E_{s,n_{y}}E^{*}_{y}
\end{bmatrix}. 
\label{eq1}
\end{equation}
Let us substitute the expression of   $\mathbf{E}^\dagger(t) =  [
\cos(\omega_{0} t) E^{\dagger}_{x}~~\sin(\omega_{0} t) E^{\dagger}_{y}
]$ with $\omega_0$ is the frequency of the probe pulse in the above equation,  
the expression of the Raman tensor reads as
\begin{equation}
\mathcal{R}_{n}(t)  = \sin[(n \omega_{\textrm{ph}}+\omega_{0})t]  
 \begin{bmatrix}
\cos(\omega_{0} t)	E_{s,n_{x}} \\
\sin(\omega_{0} t)	E_{s,n_{y}}	
\end{bmatrix}.
\label{eq2}
\end{equation} 
Here, $\omega_{\textrm{ph}}$ is the frequency of the phonon mode. 
On operating $\mathcal{D}_1$ on $\mathcal{R}_n(t)$ leads the following expression of the invariant 
$\mathcal{R}_n(t)$ as
\begin{equation}
\sin[(m\omega_{\textrm{ph}}+\omega_{0})t] 
 \begin{bmatrix}
\cos(\omega_{0}t)	E_{s,n_{x}} \\
\sin(\omega_{0}t)	E_{s,n_{y}}
\end{bmatrix}
= \sin[-(n \omega_{\textrm{ph}}+\omega_{0})t]  
\begin{bmatrix}
-\cos(\omega_{0}t)	E_{s,n_{x}} \\
-\sin(\omega_{0}t)	E_{s,n_{y}}
\end{bmatrix}.
\label{selm1}
\end{equation}
Now it is straightforward  to notice that all the sidebands exhibit nonzero
 $x$ and $y$ components.   
The ellipticity and the phase difference are estimated from the nonzero components and  
our numerical results shown in Fig.~\ref{polarization_side} are consistent with the present analysis. 
If linearly polarised probe pulse is used instead of the circularly  polarised  pulse, 
the odd- and even-order sidebands are, respectively,  
polarised perpendicular and parallel   to the polarisation of probe  
pulse in the case of graphene with  $\textsf{iLO}$ phonon mode. On the other hand, 
$\textsf{iTO}$ phonon mode leads all sidebands  polarised along the direction of the probe pulse~\cite{rana2022high}.
 
Our findings are drastically different from the recent reported work where HHG is employed to probe phonon 
dynamics in hexagonal boron nitride (h-BN). 
In Ref.~\cite{ginsberg2021optically}, it has been reported that the phonon dynamics leads the generation of the forbidden second harmonic  and enhancement of the third harmonic in few-layer h-BN. 
Moreover, coherent phonon in  h-BN leads to the attenuation of the harmonic 
spectrum with  no discrete harmonics~\cite{neufeld2022probing}. 
HHG becomes sensitive to the  carrier-envelope phase of the probe 
pulse when the pulse duration and the period of the excited phonon are similar~\cite{neufeld2022probing}. 

\section{Conclusion}  
In conclusion, we have explored the potential of high-harmonic spectroscopy in probing intertwined 
coherent-phonon electron dynamics in solids. To this end, coherent excitation of 
both in-plane  $\textsf{iLO}$ and $\textsf{iTO}$ Raman-active phonon modes in graphene are considered. 
The six-fold symmetry of the graphene reduces to two-fold dynamically due to the coherent phonon excitation.  
As a result of this symmetry alteration,   symmetry-forbidden third harmonic of circularly polarised probe pulse is 
generated. Moreover, coherent phonon leads to the generation of  the sidebands corresponding to the prominent 
harmonic peaks.  
Floquet formalism of the dynamical symmetries of the system is applied to understand the properties of the 
sidebands.  
It is found that the positions of the sidebands  are determined by the energy of the excited phonon modes.
Moreover, the dynamical symmetries of the system, consists of  graphene with an excited 
phonon mode and probe pulse, determine the polarisation of the sidebands. 
Thus, polarisation properties of the sidebands are a sensitive probe of the dynamical symmetries. 
Present study could be extended to bilayer graphene where infrared-active phonon modes can be 
expressed in terms of  double degenerate in-plane Raman-active phonon modes of monolayer graphene~\cite{gierz2015phonon, pomarico2017enhanced}. 
Moreover, this work provides a platform to study   non-linear phononics with sub-cycle temporal resolution. 

\section*{Acknowledgements}
We acknowledge fruitful discussion with M S Mrudul (Uppsala University),  
and Klaus Reimann (MBI Berlin).  
G. D. acknowledges support from Science and Engineering Research Board (SERB) India 
(Project No. MTR/2021/000138).

%

\end{document}